\documentclass[12pt]{article}
\catcode`\@=11
\@addtoreset{equation}{section}

\global\arraycolsep=2pt
\usepackage{mathrsfs,amsbsy,amssymb,latexsym,amsfonts,amsmath,cite}
\usepackage{graphicx,color}
\usepackage{physics}
\usepackage{mathtools}
\usepackage{ulem}
\usepackage{caption}
\usepackage{here,comment}
\usepackage{url}
\usepackage{hyperref}

\usepackage{mathrsfs,amsbsy,amssymb,latexsym,amsfonts,amsmath,cite,bm}
\usepackage{graphicx,color}
\usepackage{mathtools}
\usepackage{enumerate}

\DeclareFontFamily{U}{BOONDOX-calo}{\skewchar\font=45 }
\DeclareFontShape{U}{BOONDOX-calo}{m}{n}{
  <-> s*[1.05] BOONDOX-r-calo}{}
\DeclareFontShape{U}{BOONDOX-calo}{b}{n}{
  <-> s*[1.05] BOONDOX-b-calo}{}
\DeclareMathAlphabet{\mathcalboondox}{U}{BOONDOX-calo}{m}{n}
\SetMathAlphabet{\mathcalboondox}{bold}{U}{BOONDOX-calo}{b}{n}
\DeclareMathAlphabet{\mathbcalboondox}{U}{BOONDOX-calo}{b}{n}

\newcommand{\cL}{\mathcal L}

\newcommand\al[1]{\begin{align}#1\end{align}}

\newcommand\als[1]{\begin{align}\begin{split}#1\end{split}\end{align}}

\allowdisplaybreaks

\begin{document}

\renewcommand{\thefootnote}{\fnsymbol{footnote}}

\begin{flushright} 
RIKEN-iTHEMS-Report-25, STUPP-25-288
\end{flushright}
\vspace*{0.5cm}

\begin{center}
{\Large \bf  The Courant-Hilbert construction \\ 
in 4D Chern-Simons theory
}
\vspace*{2cm} \\
{\large  Osamu Fukushima\footnote{E-mail:~osamu.fukushima$\_$at$\_$riken.jp}, 
Takaki Matsumoto\footnote{E-mail:~takaki-matsumoto$\_$at$\_$ejs.seikei.ac.jp}
and Kentaroh Yoshida\footnote{E-mail:~kenyoshida$\_$at$\_$mail.saitama-u.ac.jp
}} 
\end{center}

\vspace*{0.4cm}

\begin{center}
$^{\ast}${\it iTHEMS, RIKEN, Wako, Saitama 351-0198, Japan}
\end{center}
\begin{center}
$^{\dagger}${\it Seikei University, 
3-3-1 Kichijoji-Kitamachi, Musashino-shi, Tokyo 180-8633, Japan}
\end{center}
\begin{center}
$^{\ddagger}${\it Graduate School of Science and Engineering, Saitama University, 255 Shimo-Okubo, Sakura-ku, Saitama 338-8570, Japan}
\end{center}

\vspace{1cm}

\begin{abstract}
We consider the Courant-Hilbert (CH) construction of integrable deformations of a two-dimensional principal chiral model (2D PCM) in the context of the four-dimensional Chern-Simons (4D CS) theory. According to this construction, an integrable deformation of 2D PCM is characterized by a boundary function. As a result, the master formula obtained from the 4D CS theory should be corrected by the trace of the energy-momentum tensor so as to support the CH construction. We present some examples of deformation including the $T\bar{T}$-deformation, the root $T\bar{T}$-deformation, the two-parameter mixed deformation, and a logarithmic deformation. Finally, we discuss some generalizations and potential applications of this CH construction. 
\end{abstract}

\setcounter{footnote}{0}
\setcounter{page}{0}
\thispagestyle{empty}

\newpage

\tableofcontents

\renewcommand\thefootnote{\arabic{footnote}}

\section{Introduction}\label{sec:introduction}

It is a fascinating subject to consider a unified description of integrable systems. One of the promising candidates is the four-dimensional Chern-Simons (4D CS) theory proposed by Costello and Yamazaki \cite{Costello:2019tri}, from which two-dimensional integrable models can be derived\footnote{For a nice review of 4D CS theory, see \cite{Lacroix:2021iit}.}. For example, one can derive  two-dimensional non-linear sigma models (2D NLSMs) such as a principal chiral model (PCM) \cite{Costello:2019tri,Delduc:2019whp} and symmetric coset sigma model (SCSM) \cite{Fukushima:2020dcp}. Yang-Baxter deformations\footnote{For a concise review of Yang-Baxter deformations, see a little book \cite{yoshida2021yang}.} of PCM  \cite{Klimcik:2002zj,Klimcik:2008eq,Matsumoto:2015jja}, SCSM \cite{Delduc:2013fga} and the AdS$_5\times$S$^5$ superstring \cite{Delduc:2013qra,Kawaguchi:2014qwa} can also be derived from the 4D CS theory \cite{Delduc:2019whp,Fukushima:2020dcp}.
For other related models, see~\cite{Schmidtt:2019otc,Fukushima:2020kta,Tian:2020ryu,Tian:2020pub,Fukushima:2020tqv,Lacroix:2020flf,Caudrelier:2020xtn,Fukushima:2021eni,Stedman:2021wrw,Fukushima:2021ako,Liniado:2023uoo,Berkovits:2024reg}.

\medskip 

It is interesting to consider how to describe the $T\bar{T}$-deformation \cite{Cavaglia:2016oda,Smirnov:2016lqw} and the root $T\bar{T}$-deformation \cite{Rodriguez:2021tcz,Babaei-Aghbolagh:2022uij,Ferko:2022cix,Babaei-Aghbolagh:2022leo,Tempo:2022ndz} in the context of the 4D CS theory. Both of them are known as integrable deformations which preserve the integrability of the seed theory. The integrability is preserved under the $T\bar{T}$-deformation by construction \cite{Cavaglia:2016oda, Smirnov:2016lqw}. The classical integrability under the root $T\bar{T}$-flow is shown in \cite{Borsato:2022tmu}. 
In an intriguing work by Ferko and Smith \cite{Ferko:2024ali}\footnote{For related earlier studies, see~\cite{Ivanov:2001ec,Ivanov:2002ab,Ivanov:2003uj,Ivanov:2012bq}.}, the extension of the 2D PCM with auxiliary fields was presented. 
This auxiliary field sigma model (AFSM) admits a Lax pair whose flatness condition is equivalent to the equations of motion of the model under the constraints imposed by auxiliary equations. It includes an arbitrary scalar function as well, and the choice of this function determines integrable deformations, which include the $T\bar{T}$-deformation and the root $T\bar{T}$-deformation.

\medskip

The AFSM approach was initially introduced to reformulate $T\bar{T}$-like flows and duality-invariant nonlinear electrodynamics in terms of auxiliary fields \cite{Ferko:2023wyi}, and has been further developed and generalized in various directions, including applications to T-duality \cite{Bielli:2024khq}, integrable higher-spin deformations \cite{Bielli:2024ach,Bielli:2025uiv}, (bi)-Yang-Baxter deformations \cite{Bielli:2024fnp}, deformations of (semi-)symmetric space sigma models such as $\mathbb{CP}^{N-1}$ \cite{Bielli:2024oif,Ferko:2025bhv,Cesaro:2024ipq}. These results illustrate the broad applicability of the AFSM framework and its relevance for the systematic study of integrable deformations of 2D field theories.
In the work \cite{Fukushima:2024nxm}, the 4D CS theory has been extended by including auxiliary fields and an arbitrary scalar function so as to reproduce the AFSM. Hence the $T\bar{T}$-deformation and the root $T\bar{T}$-deformation have been derived from the auxiliary field CS (AFCS) theory. 

\medskip 

An unsatisfactory point of the AFCS theory is that the auxiliary fields are included and it seems impossible to integrate them out explicitly. Towards resolving this issue, very recently, an excellent work was done by Babaei-Aghbolagh, Chen and He \cite{Babaei-Aghbolagh:2025uoz,Babaei-Aghbolagh:2025hlm}. According to these works, integrable deformations of 2D PCM, including the $T\bar{T}$-deformation and the root $T\bar{T}$-deformation, can be reproduced by solving a partial differential equation with some  boundary functions. The method to solve it was presented by Courant and Hilbert \cite{Courant}. For brevity, we shall refer to this construction as the Courant-Hilbert (CH) construction.   

\medskip 

In this paper, we consider the CH construction of integrable deformations of 2D PCM in the original 4D CS theory. As a result, we find that the master formula obtained from the 4D CS theory should be corrected by the trace of the energy-momentum tensor so as to support the CH construction. We present some examples of integrable deformations of 2D PCM including the $T\bar{T}$-deformation, the root $T\bar{T}$-deformation, the two-parameter mixed deformation, and a logarithmic deformation. Finally, we discuss some generalizations and potential applications of the CH construction discussed here. 

\medskip 

This paper is organized as follows. In section 2, we introduce how to derive an integrable sigma model from the 4D CS theory. In section 3, the CH construction of integrable deformations of 2D PCM is considered in the 4D CS theory. We show that the master formula should be corrected by the trace of the energy-momentum tensor. In section 4, we describe the CH construction of integrable deformations of 2D PCM and present some examples. Section 5 is devoted to the conclusion and discussion. Appendix A summarizes some expressions of the $T\bar{T}$-operator and the root $T\bar{T}$-operator. In Appendix B, we describe a general solution to the root $T\bar{T}$-flow equation.

\paragraph{NOTE added:} While preparing this paper, we received an interesting work \cite{Sakamoto:2025hwi}. Some parts of it are indirectly related to our results. The CH construction of integrable models has been discussed in another context, root $T\bar{T}$-unification \cite{Babaei-Aghbolagh:2025uoz,Babaei-Aghbolagh:2025hlm}.

\section{4D CS theory}

In this section, we introduce how to derive a 2D integrable sigma model (2D ISM) from the 4D CS theory \cite{Costello:2019tri} by basically following the procedure \cite{Delduc:2019whp}. Our notations and conventions are also explained here. For a nice review, see \cite{Lacroix:2021iit}.  

\subsection{From 4D CS to 2D ISM}

The 4D CS theory is a kind of gauge theory with a gauge field $A$ and a gauge group $G$\,. For simplicity, assume that the four-dimensional spacetime is given by $\mathcal{M} \times \mathbb{C}P^1$ where $\mathcal{M}$ is two-dimensional Minkowski spacetime with coordinates $x^{\mu} =(\tau,\sigma)$ and metric $\eta_{\mu\nu}={\rm diag}(-1,1)$\,. The coordinate of $\mathbb{C}P^1$ is $z$\,. Then the classical action is defined as 
\begin{align}
    S[A] \equiv \frac{i}{4\pi} \int_{\mathcal{M}\times \mathbb{C}P^1}\omega \wedge CS(A)\,. \label{CS}
\end{align}
Here $\omega$ is a $(1,0)$-form defined as 
\begin{align}
    \omega \equiv \varphi (z) dz\,,  
\end{align}
where $\varphi(z)$ is a meromorphic function defined on $\mathbb{C}P^1$\,. Since $\varphi(z)$ is complex-valued, the gauge group $G$ and the associated Lie algebra should be complexified as $G^{\mathbb{C}}$ and $\mathfrak{g}^{\mathbb{C}}$\,, respectively, to ensure the reality of the classical action. For the reality condition, see \cite{Delduc:2019whp}. Then the gauge field $A$ takes a value in $\mathfrak{g}^{\mathbb{C}}$ and the CS three-form $CS(A)$ is defined as 
\begin{align}
    CS(A) \equiv \left\langle A, dA + \frac{2}{3}A \wedge A \right\rangle\,, 
\end{align}
where the bilinear form $\langle ~, ~\rangle\,:\,\mathfrak{g}^{\mathbb{C}} \times \mathfrak{g}^{\mathbb{C}} \rightarrow \mathbb{C}$ is non-degenerate and symmetric.  

\medskip 

The action (\ref{CS}) has an extra gauge symmetry under the transformation
\begin{align}
A \rightarrow A + \chi \, dz\,,
\end{align}
because $\omega$ is a $(1,0)$-form. By using this gauge transformation, the gauge $A_z=0$ is taken. 

\medskip 

The equations of motion (eoms) are given by 
\begin{align}
   \omega \wedge F(A) &= 0\,, \qquad F(A) \equiv dA + A \wedge A\,, 
   \label{1}\\  
   d\omega \wedge \langle A, \delta A \rangle &= 0\,. \label{2}
\end{align}
The first one (\ref{1}) is called the bulk eom and the second one (\ref{2}) is the boundary eom. If $\omega$ is holomorphic, the boundary eom (\ref{2}) becomes trivial since $d\omega=0$. But since now $\omega$ is meromorphic, $d\omega$ may provide the delta-function. The boundary eom (\ref{2}) is supported just on the poles of $\varphi(z)$ and indicates that this delta function is canceled out by a zero of the gauge field $A$ through the formula $x\,\delta(x) =0$\,. Similarly, the bulk eom allows $A$ to have poles. Then $F(A)$ may have delta-function contributions, but they are canceled out by zeros of $\varphi(z)$\,. This cancellation in the bulk eom (\ref{1}) is closely related to the pole structure of the Lax form in the resulting 2D ISM. 

\medskip 

Then let us consider a formal gauge transformation, 
\begin{align}
 A = - d\hat{g} \,\hat{g}^{-1} + \hat{g}\,\mathfrak{L}\, \hat{g}^{-1}\,, 
 \label{gauge}
\end{align}
where $\hat{g}$ is a smooth group element $\hat{g}:\mathcal{M}\times \mathbb{C}P^1 \rightarrow G^{\mathbb{C}}$\,. By taking this transformation so that the $\bar{z}$-component of $A$ becomes pure gauge, the gauge condition $\mathfrak{L}_{\bar{z}}=0$ can be realized. As a result, the one-form $\mathfrak{L}$ is expressed as 
\begin{align}
 \mathfrak{L} = \mathfrak{L}_{\tau} \,d\tau + \mathfrak{L}_{\sigma} \,d\sigma\,.  
\end{align}
We shall refer to $\mathfrak{L}$ the Lax form. 
By using the expression (\ref{gauge}), the bulk eom becomes 
\begin{align}
    \partial_{\tau}\mathfrak{L}_{\sigma} -  \partial_{\sigma}\mathfrak{L}_{\tau} + [\mathfrak{L}_{\tau},\mathfrak{L}_{\sigma}] &= 0\,, \label{bulk1}\\ 
    \omega \wedge \partial_{\bar{z}} \mathfrak{L} &= 0\,. \label{bulk2}
\end{align}
The explicit form of $\mathfrak{L}$ should be determined so as to satisfy the conditions (\ref{bulk1}) and (\ref{bulk2})\,. In particular, the zero structure of $\omega$ is closely related to the pole structure of $\mathfrak{L}$\,. A general ansatz \cite{Delduc:2019whp} is given by 
\begin{align}
  \mathfrak{L} = \sum_{i=1}^{\sharp\mathfrak{z}_{(+)}}\frac{V_{+}^{(i)}}{z-z_i}\,d\sigma^+ 
  + \sum_{j=1}^{\sharp\mathfrak{z}_{(-)}}\frac{V_{-}^{(j)}}{z-z_j}\,d\sigma^- 
+ U_{\tau}\,d\tau + U_{\sigma}\, d\sigma 
  \,, \label{general}
\end{align}
where $V_{+}^{(i)}(\tau,\sigma)$\,, $V_{-}^{(j)}(\tau,\sigma)$\,, $U_{\tau}(\tau,\sigma)$ and $U_{\sigma}(\tau,\sigma)$ are smooth functions on $\mathcal{M}$ and are independent of $z$\,. The light-cone coordinates are defined as 
\begin{align}
    \sigma^{\pm} = \frac{1}{2}(\tau \pm \sigma)\,. 
\end{align}
The set of the zeros of $\varphi$\,, $\mathfrak{z}$ is decomposed into the two sets as follows:
\begin{align}
    \mathfrak{z} = \mathfrak{z}_{(+)} \cup \mathfrak{z}_{(-)}\,.
\end{align}
This decomposition depends on the 2D ISM we want to obtain. 

\medskip 

Then, by performing the complex integral \cite{Benini:2020skc}, the classical action of the 4D CS theory is dimensionally reduced to the following 2D form \cite{Delduc:2019whp}: 
\begin{align}
S[\{g_x\}_{x\in \mathfrak{p}}] = \frac{1}{2}\sum_{x\in \mathfrak{p}}\int_{\mathcal{M}} \left\langle {\rm res}_x(\varphi \mathfrak{L}), g_x^{-1}dg_x
\right\rangle + \frac{1}{2}\sum_{x\in \mathfrak{p}} ({\rm res}_{x}\varphi)\int_{\mathcal{M}\times I} I_{\rm WZ}[g_x]\,,
\label{master-formula}
\end{align}
where $\mathfrak{p}$ is the set of poles of $\varphi$\,, $g_x \equiv \hat{g}|_{\mathcal{M} \times \{x\}}$ and $I_{\rm WZ}$ is the standard Wess-Zumino term defined as 
\begin{align}
 I_{\rm WZ}[u] \equiv \frac{1}{3}\langle u^{-1}du, u^{-1}du\wedge u^{-1}du
 \rangle\,.  
\end{align}
This is the master formula to provide the resulting action of 2D ISM. 

\medskip 

Finally, the recipe to derive a 2D ISM from the 4D CS theory is the following: 
\begin{enumerate}
    \item Choose a meromorphic function $\varphi$\,.
    \item Construct a Lax form from the zeros of $\varphi$ consistently with the bulk eom. 
    \item Impose boundary conditions of $A$ at the poles of $\varphi$ so as to satisfy the boundary eom.
    \item Evaluate the master formula and obtain the resulting 2D ISM. 
\end{enumerate}

\subsection{Example: Derivation of 2D PCM with the Wess-Zumino term}

As a concrete example, let us here derive 2D PCM with the Wess-Zumino (WZ) term from the 4D CS theory. This would be a good exercise to understand the punchline of section 3. This part follows the work \cite{Delduc:2019whp}. 

\medskip 

To derive 2D PCM with the WZ term, we take the following meromorphic function: 
\als
{
	\omega = \varphi(z)dz \equiv K \frac{1-z^2}{(z-k)^2}dz\,,
    \label{mero-PCM}
}
where $K$ and $k$ are real constants. 
The set of poles is $\mathfrak{p}= \{k,\infty\}$ and the one of zeros is $\mathfrak{z}=\{1,-1\}$\,. Then the boundary conditions for $A$ are simply taken as  
\begin{align}
 A|_{z=k}=A|_{z=\infty}=0\,. \label{bc-A} 
\end{align}
After performing the complex integral, the group element $\hat{g}$ is restricted as 
\begin{align}
\hat{g}|_{z=k} =g_k = g(\tau,\sigma)\,, \qquad 
\hat{g}|_{z=\infty} = g_{\infty} = 1\,. 
\end{align}
The group element $g_{\infty}$ has been set to 1 by using the residual gauge symmetry \cite{Delduc:2019whp}. Then the boundary conditions in (\ref{bc-A}) can be rewritten as 
\begin{align}
  & A|_{z=k} =  -dg g^{-1} + g \mathfrak{L} g^{-1} = 0\,, \label{bc1} \\
  & A|_{z=\infty} = \mathfrak{L}|_{z=\infty} = 0\,. \label{bc2} 
\end{align}

The ansatz for the Lax form is given by 
\begin{align}
    \mathfrak{L} = \frac{V_{+}^{(1)}(\tau,\sigma)}{z-1}\,d\sigma^+ + \frac{V_{-}^{(1)}(\tau,\sigma)}{z+1}\,d\sigma^- + U_{\tau}(\tau,\sigma)\, d\tau  + U_{\sigma}(\tau,\sigma)\, d\sigma\,. \label{Lax-PCM-ansatz}
\end{align}
The assignment rule of zeros led to the trivial index $(1)$ and hence we will omit it hereafter. 
The condition (\ref{bc2})\,, $U_{\tau} = U_{\sigma} =0$\,. Then, the condition (\ref{bc1}) leads to 
\begin{align}
  V_{\pm} = (k \mp 1)\,j_{\pm}\,, \qquad j_{\pm} \equiv  g^{-1}\partial_{\pm}g\,.  
\end{align}
Thus the Lax form is determined as 
\begin{align}
  \mathfrak{L} = \frac{k-1}{z-1}\,j_+  d\sigma^+ + \frac{k+1}{z+1}\, j_- d\sigma^-\,.   
\end{align}
The resulting action is given by 
\begin{align}
  S[g] = \frac{K}{2} \int_{\mathcal{M}} d\sigma \wedge d\tau\, \langle j_+, j_- \rangle  + Kk I_{\rm WZ}[g]\,.
\end{align}
This is nothing but the classical action of 2D PCM with the WZ term.

\section{Deforming 2D PCM}

In this section, let us consider integrable deformations of 2D PCM without the WZ term, for simplicity. We use the following meromorphic one-form:
\als
{
	\omega = \varphi(z)dz \equiv \frac{1-z^2}{z^2}dz\,,
    \label{mero-PCM-simple}
}
which is obtained from (\ref{mero-PCM}) by setting $k=0$ and $K=1$\,. The boundary conditions are the same as (\ref{bc-A})\,. This setup corresponds to taking 2D PCM as the seed theory. 

\medskip 

From the expression of $\varphi$\,, the 2D action given by 
\als
{
	\label{action}
	S
	=
	\frac{1}{2}\int_{\mathcal{M}} {\rm tr} \left(
	\mathrm{res}_{z=0}\,(\varphi\,\mathfrak{L})\wedge g^{-1}dg \right) \,,
}
Here the bracket $\langle~,~\rangle$ has been replaced by the trace operation, ``${\rm tr}$'' by assuming a finite matrix representation.  

\medskip 

So far, there is no difference from the computation in section 2.2. Here, instead of the usual Lax ansatz (\ref{Lax-PCM-ansatz})\,, let us try to examine the following ansatz: 
\als
{
    \mathfrak{L}
    &= \frac{V_{+}(\tau,\sigma) + z\, K_{+}(\tau,\sigma)}{z^2-1}\,d\sigma^+ + \frac{V_{-}(\tau,\sigma) - z\, K_{-}(\tau,\sigma)}{1-z^2}\,d\sigma^- \\
    & \quad  + U_{\tau}(\tau,\sigma) d\tau  + U_{\sigma}(\tau,\sigma) d\sigma\,. 
    \label{extend}
}
Here $K_{\pm}(\tau,\sigma)$ are smooth functions on $\mathcal{M}$ newly introduced here. Although the zero assignment rule is also changed\footnote{Note here that if the sets $\mathfrak{z}_{(+)}$ and $\mathfrak{z}_{(-)}$ have a nontrivial intersection as in the present Lax ansatz, the term proportional to $\varphi(z)\langle \mathfrak{L}, d\mathfrak{L}\rangle$ can remain in \eqref{master-formula} after the reduction to the 2D action in general. However, this term vanishes in the present case. 
}, this ansatz (\ref{extend}) is reduced to (\ref{Lax-PCM-ansatz}) when $K_\pm = V_\pm$\,. From the boundary condition (\ref{bc2})\,, $U_{\tau}$ and $U_{\sigma}$ should vanish again. Since the boundary condition (\ref{bc1}) at $z=0$ is not changed, we obtain again that 
\begin{align}
 V_{\pm} = \mp j_{\pm}\,, \qquad j_{\pm} =  g^{-1}\partial_{\pm}g\,.  
\end{align}
By introducing the following definition, 
\begin{align}
K_{\pm} \equiv \mp \mathfrak{J}_{\pm}\,, 
\end{align}
the components of the resulting Lax form are given by  
\als
{
	\label{lax-ansatz}
	\mathfrak{L}_\pm
	=
	\frac{j_\pm \pm z \,\mathfrak{J}_\pm}{1-z^2}\,, \qquad j_{\pm}=g^{-1}\partial_{\pm} g\,.
}
Note here that the functions $\mathfrak{J}_{\pm}$ have not been determined and remain arbitrary. 

\medskip 

Let us examine the eoms obtained from the 4D CS theory. First of all, the boundary eom is obviously satisfied under the boundary condition (\ref{bc-A})\,. The remaining is to examine the bulk eoms (\ref{bulk1}) and (\ref{bulk2})\,. For the bulk eom (\ref{bulk2})\,, it is helpful to notice that the following relation holds for an arbitrary holomorphic function $f(z)$: 
\als
{
	\varphi(z)\,\partial_{\bar z}\frac{f(z)}{1-z^2}
	&=
	\frac{1-z^2}{z^2}\frac{f(z)}{1-z}\partial_{\bar z}\frac{1}{1+z} +
	\frac{1-z^2}{z^2}\frac{f(z)}{1+z}\partial_{\bar z}\frac{1}{1-z}
	\\ 
	& \propto \, 
	\frac{1+z}{z^2}f(z)\,\delta^{(2)}(1+z)
	+ \frac{1-z}{z^2}f(z)\,\delta^{(2)}(1-z)\\
	&=0\,, 
}
where $\varphi$ is defined in \eqref{mero-PCM-simple}\,. Thus the second bulk eom (\ref{bulk2}) is also obviously satisfied. 

\medskip 

The first bulk eom (\ref{bulk1}) imposes some conditions on $\mathfrak{J}_\mu$\,. By substituting the ansatz \eqref{lax-ansatz} into \eqref{bulk1}\,, the bulk eom (\ref{bulk1}) is evaluated as follows:
\als
{
	\label{be1second}
	0
	& = \partial_{[+}\mathfrak{L}_{-]} + [\mathfrak{L}_+, \mathfrak{L}_-]\\
	& =
	\frac{\partial_{[+} j_{-]} - z(\partial_+\mathfrak{J}_- + \partial_-\mathfrak{J}_+)}{1-z^2}
	+ \frac{ [j_+,j_-] + z([\mathfrak{J}_+,j_-] - [j_+,\mathfrak{J}_-])
	- z^2[\mathfrak{J}_+,\mathfrak{J}_-]}{(1-z^2)^2}\\
	& =
	- \frac{z(\partial_+\mathfrak{J}_- + \partial_-\mathfrak{J}_+)}{1-z^2}
	+ \frac{ z([\mathfrak{J}_+,j_-] - [j_+,\mathfrak{J}_-]) -
	z^2([\mathfrak{J}_+,\mathfrak{J}_-] - [j_+,j_-])}{(1-z^2)^2}\,.
}
Here in the second equality we have used the flatness of $j_\mu$\,, 
\begin{align}
    \partial_{\mu}j_{\nu} - \partial_{\nu} j_{\mu} + [j_\mu,j_\nu] = 0\,. 
\end{align} 
The last expression can be expanded in terms of $z$ as 
\als
{
    & (-(\partial_+\mathfrak{J}_- + \partial_-\mathfrak{J}_+) +[\mathfrak{J}_+,j_-] - [j_+,\mathfrak{J}_-]) z -([\mathfrak{J}_+,\mathfrak{J}_-] - [j_+,j_-])z^2\\ 
    & + (-(\partial_+\mathfrak{J}_- + \partial_-\mathfrak{J}_+) + 2([\mathfrak{J}_+,j_-] - [j_+,\mathfrak{J}_-]))z^3 + \mathcal{O}(z^4)\,. 
}
One can easily check that all of the coefficients vanish under the following conditions: 
\al
{
	& \label{2deom}
	(\partial_+\mathfrak{J}_- + \partial_-\mathfrak{J}_+) = -2 \partial^\mu \mathfrak{J}_\mu
	= 0\,, \\ 
	\label{com1}
    &
	[\mathfrak{J}_+,j_-] = [j_+,\mathfrak{J}_-]\,,\\
	\label{com2}
	&[\mathfrak{J}_+,\mathfrak{J}_-] = [j_+,j_-]\,.
}
Thus, these are non-trivial conditions that $\mathfrak{J}$ must satisfy. Note here that, based on the standard knowledge about the Lax pair, the first (\ref{2deom}) should be the ``on-shell'' condition, while the others (\ref{com1}) and (\ref{com2}) are off-shell ones. This observation will be significant later. 

\medskip 

The next is to find out the explicit form of $\mathfrak{J}$ that satisfies the conditions (\ref{2deom})-(\ref{com2}). For this purpose, we introduce two functionally independent Lorentz scalars defined as 
\als
{
    \label{x-variable}
	& x_1 \equiv \mathrm{tr}\,j^\mu j_\mu = - \mathrm{tr}\,j_+j_-,\\
	& x_2 \equiv \mathrm{tr}\,j^\mu j_\nu\, \mathrm{tr}\,j^\nu j_\mu
	= \frac{1}{2} \mathrm{tr}\,j_+j_+\,\mathrm{tr}\,j_-j_- + \frac{1}{2}(\mathrm{tr}\,j_+j_-)^2\,.
}
As described in \cite{Ferko:2022cix}, all other Lorentz scalars constructed from $\mathrm{tr}\,j^\mu j_\nu$ can be expressed as a polynomial of $x_1$ and $x_2$ by means of trace identities for spacetime $2\times 2$ matrices such as
\als
{
	&
	\label{trace-id}
	\mathrm{tr}\,j^\mu j_\nu\, \mathrm{tr}\,j^\nu j_\rho\, \mathrm{tr}\,j^\rho j_\mu
	= \frac{3}{2}x_1x_2 - \frac{1}{2}x_1^3\,,\\
	&
	\mathrm{tr}\,j^\mu j_\nu\, \mathrm{tr}\,j^\nu j_\rho\,
	\mathrm{tr}\,j^\rho j_\sigma\, \mathrm{tr}\,j^\sigma j_\mu 
	= x_1^2x_2 - \frac{1}{2}x_1^4 + \frac{1}{2}x_2^2\,.
}
Hence, by using $x_1$ and $x_2$\,, let us express $\mathfrak{J_\mu}$ as 
\als
{
	\label{object}
	\mathfrak{J}_\mu
	= 2f_1(x_1,x_2)\, j_{\mu} + 4f_2(x_1,x_2)\, j_{\nu}\, \mathrm{tr}\, j^{\nu}j_{\mu}\,, 
}
where $f_1$ and $f_2$ are arbitrary functions of $x_1$ and $x_2$ at this moment, and the numerical coefficients are set for later convenience. In the light-cone coordinates, we have 
\als
{
	&
	\mathfrak{J}_+
	= 2(f_1 + x_1 f_2) j_+ - 2f_2\, j_- \,\mathrm{tr}\, j_+ j_+\,,\\
	&
	\mathfrak{J}_-
	= 2(f_1 + x_1 f_2) j_- - 2f_2\, j_+ \,\mathrm{tr}\, j_- j_-\,.
}
One might expect that additional terms, such as $f_3(x_1,x_2)j_{\rho}\, \mathrm{tr}\, j^{\rho}j_{\nu}\,\mathrm{tr}\, j^{\nu}j_{\mu}$, should also be included. However, such terms reduce to Lorentz scalars constructed from $\mathrm{tr}\,j^\mu j_\nu$ in the action \eqref{action}, and merely redefine $f_1$ or $f_2$ via trace identities. Thus, the two terms in \eqref{object} are sufficient.

\medskip

The ansatz \eqref{object} obviously satisfies the first off-shell condition (\ref{com1}). The second off-shell condition (\ref{com2}) is also satisfied if and only if $f_1$ and $f_2$ obey
\als
{
	\label{condition-for-f}
	4(f_1+x_1f_2)^2 - 4(2x_2-x_1^2)f_2^2 = 1\,.
}
The remaining task is to check the condition (\ref{2deom})\,. This condition is somewhat different from the others, because it should be the ``on-shell'' condition from the standard understanding of the flatness condition for the Lax pair. Hence, the condition (\ref{2deom}) should be checked by the on-shell analysis. To this end, let us substitute the Lax ansatz (\ref{lax-ansatz}) with \eqref{object} into the action (\ref{action}) and study its variation.

\medskip

For simplicity, we will focus upon the Lagrangian density rather than the action. The Lagrangian in (\ref{action}) is evaluated as 
\al
{
	\label{lagrangian}
	\mathcal{L}
	&=
	\frac{1}{2}\mathrm{tr}\,\mathfrak{J}^\mu j_\mu\\
	\label{lagrangian2}
    &=
	x_1f_1 + 2x_2 f_2\,. 
}
By taking a variation of (\ref{lagrangian}) with respect to $g$\,, we obtain
\als
{
	\label{variation}
	\delta\mathcal{L}
	=
	\frac{1}{2}\mathrm{tr}\, \delta \mathfrak{J}^\mu\,j_\mu
	+ \frac{1}{2}\mathrm{tr}\, \mathfrak{J}^\mu\,\delta j_\mu\,.
}
The first term can be written as
\als
{
	\label{variation-1st}
	\frac{1}{2}\mathrm{tr}\, \delta\mathfrak{J}^\mu\, j_\mu
	&= x_1\delta f_1 + 2x_2\delta f_2 + \frac{1}{2}f_1\delta x_1 + \frac{3}{2}f_2\delta x_2 \\
	&= \delta(x_1f_1 + 2x_2 f_2) - \frac{1}{2}f_1\delta x_1 - \frac{1}{2}f_2\delta x_2
}
where in the first equality we have used
\als
{
	\mathrm{tr}\,(\delta j^\mu) j_\mu = \frac{1}{2}\delta x_1\,,\qquad
	\mathrm{tr}\,(\delta (j_\nu \mathrm{tr}\,j^\nu j^\mu)) j_\mu = \frac{3}{4}\delta x_2\,,
}
and the second equality follows from
\als
{
	x_1\delta f_1 + 2x_2\delta f_2
	&=
	\delta(x_1f_1 + 2x_2 f_2) - f_1\delta x_1 - 2f_2\delta x_2\,.
}

\medskip

We next focus on the second term in \eqref{variation}. By evaluating the variation of $j^{\mu}$\,,
\als
{
	\delta j^\mu
	=
	(\delta g^{-1})\partial_\mu g + g^{-1}\partial_\mu\delta g
	=
	-g^{-1}\delta g j_\mu + g^{-1}\partial_\mu\delta g\,,
}
the second term in (\ref{variation}) can be rewritten as
\als
{
	\frac{1}{2}\,\mathrm{tr}\,\mathfrak{J}^\mu\,\delta j_\mu
	&=
	- \frac{1}{2}\,\mathrm{tr}\,\mathfrak{J}^\mu\,g^{-1}\delta g\,j_\mu
	+ \frac{1}{2}\,\mathrm{tr}\,\mathfrak{J}^\mu\, g^{-1}\partial_\mu\delta g\\
	&=
	- \frac{1}{2}\,\mathrm{tr}\,j^\mu\mathfrak{J}_\mu\,g^{-1}\delta g
	-\frac{1}{2}\,\mathrm{tr}\,\mathfrak{J}^\mu\,(\partial_\mu g^{-1})\delta g
	-\frac{1}{2}\,\mathrm{tr}\,(\partial^\mu\mathfrak{J}_\mu)\,g^{-1}\delta g\\
	&=
	-\frac{1}{2}\,\mathrm{tr}\,(\partial^\mu\mathfrak{J}_\mu)\,g^{-1}\delta g\,,
}
where we have dropped total derivative terms, and the condition $[\mathfrak{J}^\mu, j_\mu]=0$, which is equivalent to \eqref{com1}, has been used in the third equality.

\medskip

In summary, we have found that the variation of the Lagrangian \eqref{lagrangian} is given by 
\als
{
	\label{variation-final}
	\delta\mathcal{L}
	=
	\delta(x_1f_1 + 2x_2 f_2)
	- \frac{1}{2}(f_1\delta x_1 + f_2\delta x_2)
	- \frac{1}{2}\,\mathrm{tr}\,(\partial^\mu\mathfrak{J}_\mu)\,g^{-1}\delta g\,,
}
up to total derivative terms. In order for (\ref{2deom}) to hold as the on-shell condition, let us take the following forms of $f_1$ and $f_2$\,, respectively, 
\als
{
	f_1 = \partial_1\mathcal{F}(x_1,x_2)\,,\qquad
	f_2 = \partial_2\mathcal{F}(x_1,x_2)\,.
}
Here $\mathcal{F}$ is a function of $x_1$ and $x_2$ obeying
\als
{
	\label{universal-PDE}
	4(\partial_1\mathcal{F}+x_1\partial_2\mathcal{F})^2 -
	4(2x_2-x_1^2)(\partial_2\mathcal{F})^2 = 1\,,
}
which follows from the condition \eqref{condition-for-f}. With this choice, $\delta\mathcal{L}$ becomes 
\als
{
    \label{variation-final2}
	&
	\delta\mathcal{L}
	= \delta\biggl(x_1 \partial_1\mathcal{F}
	+ 2x_2 \partial_2\mathcal{F} - \frac{1}{2}\mathcal{F}\biggr)
	- \frac{1}{2}\,\mathrm{tr}\,(\partial^\mu\mathfrak{J}_\mu)\,g^{-1}\delta g\,,
}
respectively. This suggests that the Lagrangian should be corrected as
\begin{align}
\frac{1}{2}\hat{\mathcal{L}} \equiv \mathcal{L} - \left(x_1\partial_1\mathcal{F}
	+ 2x_2 \partial_2\mathcal{F} - \frac{1}{2}\mathcal{F} \right)\, 
    \label{mod}
\end{align}
where the factor 1/2 on the left-hand side has been introduced for later convenience. Then the variation of the corrected Lagrangian $\hat{\mathcal{L}}$ leads to 
\begin{align}
  \delta \hat{\mathcal{L}}
  = - \mathrm{tr}\,\partial^\mu\mathfrak{J}_\mu\,g^{-1}\delta g\,,
\end{align}
and the eom (\ref{2deom}) is obtained from $\hat{\mathcal{L}}$ rather than $\mathcal{L}$\,. 

\medskip 

It is significant to rewrite the corrected Lagrangian (\ref{mod})\,. There are two expressions of it. The one is to just simplify the Lagrangian (\ref{mod}) as  
\begin{align}
    \hat{\mathcal{L}} = \mathcal{F}\,. 
\end{align}
Inversely speaking, the undetermined function $\mathcal{F}$ itself is the true Lagrangian. This expression can be obtained by comparing the variation (\ref{variation-final2}) with the variation of the expression (\ref{lagrangian2})\,. 
The other one is obtained by using (\ref{lagrangian2})\,, and $\hat{\mathcal{L}}$ can be expressed as 
\begin{align}
\hat{\mathcal{L}} &= \mathcal{L} 
-  \left(x_1\partial_1\mathcal{F}
	+ 2x_2 \partial_2\mathcal{F} - \mathcal{F} \right) \nonumber \\   
   &= \mathcal{L} + \frac{1}{2}\hat{T}_{~~\mu}^{\mu} \,,
    \label{mod2}
\end{align}
where after rewriting, both sides have been multiplied by 2 and in the last equality, we have used the relation (\ref{trace-EMtensor}) computed in the Appendix \ref{appB}. As a result, the original master formula has been corrected by the trace of the energy-momentum tensor for $\hat{\mathcal{L}}$~\footnote{The determinant term, which breaks classical scale invariance, derived from the relation between curved space and $T\bar{T}$-deformation \cite{Sakamoto:2025hwi} corresponds to a special case of this term here.}. 

\medskip 

Some comments for this result should be made here. First, the corrected Lagrangian $\hat{\mathcal{L}}$ itself is an undetermined function $\mathcal{F}$ which should be a solution to the PDE (\ref{universal-PDE})\,. Hence, by solving this PDE somehow, the Lagrangian is determined. This may sound weird, but this is rather better to consider integrable deformations of 2D PCM in a unified way as described in section 4. Second, the trace of the energy-momentum tensor in the corrected Lagrangian (\ref{mod2}) vanishes when the deformation is marginal like the root $T\bar{T}$-deformation and the classical scale invariance is preserved. Then the Lagrangian $\hat{\cL}$ coincides with $\mathcal{L}$ and the original master formula is valid as it is. Inversely speaking, when the deformation like the $T\bar{T}$-deformation breaks the scale invariance, the corrected term should appear inevitably. Finally, the master formula Lagrangian $\mathcal{L}$ did not work as the usual Lagrangian. This is just because the Lax form obviously depends on the Lagrangian. The insertion of the Lax form into the 4D CS action makes the definition of the ``Lagrangian'' obscure. Only the reliable criterion to declare what is the true Lagrangian is that a variation of that quantity gives rise to a correct eom. 

\medskip 

In summary, by generalizing the Lax ansatz so as to deform 2D PCM, we have obtained the corrected Lagrangian
\als
{
    \label{final-lagrangian}
    \hat{\mathcal{L}} &= \mathcal{L} + \frac{1}{2}\hat{T}_{~~\mu}^{\mu} =\mathcal{F}\,. 
}
Here the energy-momentum tensor is given by 
\als
{
    \label{em-tensor}
	\hat{T}_{\mu\nu}
	&=
	-2\frac{\partial\hat{\mathcal{L}}}{\partial \eta^{\mu\nu}} + \eta_{\mu\nu}\hat{\mathcal{L}}\,,
} 
and $\mathcal{F}$ should satisfy the PDE 
\als
{\label{constraint}
    4(\partial_1\mathcal{F}+x_1\partial_2\mathcal{F})^2 - 4(2x_2-x_1^2)(\partial_2\mathcal{F})^2 = 1\,. 
}
After deriving the explicit form of $\mathcal{F}$\,, 
the associated Lax pair is also determined by
\als
{
    \label{lax-pair}
    &\mathfrak{L}_\pm
    = \frac{j_\pm \pm z \mathfrak{J}_\pm}{1-z^2}\,, \\
    &\mathfrak{J}_\mu
    = 2(\partial_1\mathcal{F})j_\mu + 4 (\partial_2\mathcal{F})j^\nu\,\mathrm{tr}\,j_\mu j_\nu\,.
}
Hereafter, we will refer to the equations listed above though they have already appeared in the previous discussion. 

\medskip 

In the next section, we will describe how to determine $\mathcal{F}$ by following the method found by Courant and Hilbert \cite{Courant}.

\section{The Courant-Hilbert construction}

Now the resulting 2D Lagrangian is given in (\ref{final-lagrangian})\,. Still, we have to find out a possible form of $\mathcal{F}$ by solving the PDE (\ref{constraint})\,. For this problem, there is a systematic way found by Courant and Hilbert \cite{Courant}. In the following, we will describe it. 

\medskip 

First of all, let us introduce new variables defined as 
\als
{
    \label{uv-variable}
	&
	u \equiv \frac{1}{4}({\textstyle\sqrt{2x_2 - x_1^2}} - x_1)\,, \qquad 
	v \equiv \frac{1}{4}({\textstyle\sqrt{2x_2 - x_1^2}} + x_1)\,,
}
by following the work \cite{Babaei-Aghbolagh:2025uoz}. Then we can see that the inverse relations are given by
\als
{
	&
	x_1 = 2(v - u)\,, \qquad 
	x_2 = 4 (u^2 + v^2)
}
and the derivatives are related through 
\als
{
	&
	\partial_1 = \frac{1}{2(u+v)}(-v\,\partial_u + u\,\partial_v)\,, \qquad 
	\partial_2 = \frac{1}{8(u+v)}(\partial_u + \partial_v)\,.
}
In terms of $u$ and $v$, the PDE \eqref{constraint} takes the simple form
\als
{
	\label{universal-PDE2}
	\partial_u\mathcal{F}\, \partial_v\mathcal{F} = - 1\,. 
}
Deformations of 2D PCM can be studied by solving this PDE \eqref{universal-PDE2} as shown in \cite{Babaei-Aghbolagh:2025uoz}.

\medskip 

The general solution to the PDE (\ref{universal-PDE2}) has been constructed by Courant and Hilbert \cite{Courant}. The solution to \eqref{universal-PDE2} is given by
\als
{
	\label{CH-solution}
	&\mathcal{F}(u,v) = \ell(\tau) - \frac{2u}{\ell'(\tau)}\,, \qquad 
    \tau = v + \frac{u}{\ell'(\tau)^2}\,.
}
Here a new function $\ell$ is introduced through the boundary condition,  
\begin{align}
   \mathcal{F} (0,v) = \ell (v)  
\end{align}
and $\ell'=d\ell/d\tau$\,. Now the problem of finding out deformations of 2D PCM is simplified to picking up possible forms of $\ell(\tau)$\,. Several examples are found in \cite{Babaei-Aghbolagh:2025hlm,Babaei-Aghbolagh:2025uoz}. In the following, we shall list five examples of $\ell$ associated with integrable deformations. All of them have been obtained originally in \cite{Babaei-Aghbolagh:2025uoz}.  

\subsubsection*{(1) the undeformed PCM}

The first example is the simplest solution given by \cite{Russo:2024llm}
\begin{align} 
\ell(\tau) = \tau\,. 
\end{align}
Then $\ell' = 1$ and the resulting Lagrangian becomes 
\begin{align}
   \mathcal{F} = v-u = \frac{1}{2}x_1 = - \frac{1}{2} {\rm tr} (j_+ j_-)\,.
\end{align}
This is nothing but the original PCM. 

\subsubsection*{(2) the root $T\bar{T}$-deformation of PCM}

The second is a bit non-trivial solution \cite{Russo:2024llm}\footnote{A more general form is possible to satisfy the root $T\bar{T}$-flow equation. For the detail, see Appendix \ref{app:root-TT}},
\begin{align} 
\label{rootTTbar-PCM}
\ell(\tau) = {\rm e}^{\gamma} \tau\,,
\end{align}
where $\gamma$ is a real constant. Then $\ell' = {\rm e}^{\gamma}$ and the resulting Lagrangian is given by 
\begin{align}
   \mathcal{F} &= {\rm e}^{\gamma} v - {\rm e}^{-\gamma} u  \nonumber \\ 
   &= \frac{1}{2}\cosh\gamma \cdot x_1 + \frac{1}{2}\sinh\gamma \cdot {\textstyle\sqrt{2x_2 -x_1^2}}\,. 
\end{align}
This is nothing but the root $T\bar{T}$-deformed PCM \cite{Borsato:2022tmu}. 

\medskip 

In this case, the trace of the energy-momentum tensor vanishes,  
\begin{align}
   \hat{T}^{\mu}_{~~\mu}  = 0
\end{align}
because the root $T\bar{T}$-deformation is marginal.

\subsubsection*{(3) the $T\bar{T}$-deformation of PCM} 

The third one is the following solution \cite{Russo:2024llm}, 
\begin{align}
    \label{TTbar-PCM}
   \ell(\tau) = - \frac{1}{\lambda}(1-\sqrt{1+ 2\lambda \tau})\,,
\end{align}
where $\lambda$ is a real constant. 
Then the Lagrangian is evaluated as 
\begin{align}
    \mathcal{F} &= -\frac{1}{\lambda} (1 -\sqrt{1+2\lambda\tau}) - 2u\sqrt{1+2\lambda \tau} \nonumber \\ 
    &= -\frac{1}{\lambda} + \frac{1}{\lambda}\sqrt{(1 - 2\lambda u)(1 +2\lambda v) } \nonumber \\ 
    &= -\frac{1}{\lambda} + \frac{1}{\lambda}\sqrt{1 + \lambda x_1 +\frac{\lambda^2}{2}(x_1^2-x_2)}\,. 
\end{align}
This precisely agrees with the $T\bar{T}$-deformed Lagrangian computed in \cite{Chen:2021aid}. 

\medskip 

One can see that the trace of the energy-momentum tensor is expressed as  
\begin{align}
\frac{1}{2}\hat T^{\mu}_{~~\mu} = -\lambda\partial_\lambda \mathcal{F}\,.
\end{align}
For the detail of the energy-momentum tensor, see Appendix \ref{appB}.

\medskip

In particular, for this solution, one can show the following relation:
\als
{
    \label{trace-determinant}
	\frac{1}{2}\hat T^\mu{}_\mu 
    =
    - \lambda \det \hat T_{\mu\nu}\,.
}
That is, the corrected term is the determinant of $\hat{T}_{\mu\nu}$ 
as obtained in \cite{Sakamoto:2025hwi}.

\subsubsection*{(4) the two-parameter mixed deformation of PCM} 

It is known that the $T\bar{T}$-deformation commutes with the root $T\bar{T}$-deformation \cite{Ferko:2022cix}. Hence, it is possible to consider a two-parameter mixed deformation. This deformation is given by \cite{Russo:2024llm}\footnote{This $\gamma$-dependence can also be explained by the scaling argument in Appendix \ref{app:root-TT}.}
\begin{align}
   \ell(\tau) = - \frac{1}{\lambda} (1-\sqrt{1+ 2\lambda \mathrm{e}^\gamma \tau})\,, \label{4.16}
\end{align}
where $\gamma$ and $\lambda$ are the parameters for the root $T\bar{T}$-deformation and the $T\bar{T}$-deformation, respectively. Then, the function $\mathcal{F}$ is obtained as  
\begin{align}
    \mathcal{F} &= - \frac{1}{\lambda} (1 -\sqrt{1+2\lambda\mathrm{e}^\gamma\tau}) - 2u {\rm e}^{-\gamma}\sqrt{1+2\lambda\mathrm{e}^\gamma \tau} \nonumber \\ 
    &= -\frac{1}{\lambda} + \frac{1}{\lambda}\sqrt{(1 - 2 {\rm e}^{-\gamma}\lambda u)(1 +2{\rm e}^{\gamma}\lambda v) } \nonumber \\ 
    &= -\frac{1}{\lambda} + \frac{1}{\lambda}\sqrt{1 + \lambda (x_1 \cosh\gamma + \sqrt{2x_2-x_1^2}\,\sinh\gamma )  + \frac{\lambda^2}{2} (x_1^2-x_2) }\,. 
\end{align}
In this case, we can see again that 
\begin{align}
\frac{1}{2}\hat T^{\mu}_{~~\mu} = -\lambda\partial_\lambda \mathcal{F}\,.
\end{align}

\subsubsection*{(5) a logarithmic deformation of PCM} 

Finally, let us consider a logarithmic function \cite{Russo:2024llm} given by 
\begin{align}
\ell(\tau) = -\frac{1}{\lambda} \log (1-\lambda \tau)\,,
\end{align}
where $\lambda$ is a constant parameter. Then $\ell' = 1/(1-\lambda \tau)$ and we can derive 
\begin{align}
  1-\lambda \tau = \frac{-1 + \sqrt{1 + 4\lambda u (1-\lambda v)}}{2\lambda u}\,.
\end{align}
By using this relation, the function $\mathcal{F}$ is given by 
\begin{align}
    \mathcal{F} = -\frac{1}{\lambda}\log\left(
    \frac{-1 + \sqrt{1 + 4\lambda u (1-\lambda v)}}{2\lambda u}
    \right) + \frac{1}{\lambda} -\frac{1}{\lambda} \sqrt{1 + 4\lambda u (1-\lambda v)}\,.  
\end{align}
In this case, the trace of the energy-momentum tensor satisfies 
\begin{align}
    \frac{1}{2}\hat T^{\mu}_{~~\mu} = -\lambda \partial_{\lambda}\mathcal{F}\,.  
\end{align}

\section{Conclusion and Discussion}\label{sec:conclusion}

In this paper, we have studied the CH construction of integrable deformations of 2D PCM in the 4D CS theory. As a result, the master formula has been corrected by the trace of the energy-momentum tensor so as to support the CH construction. We have presented some examples of integrable deformations of 2D PCM including the $T\bar{T}$-deformation, the root $T\bar{T}$-deformation, the two-parameter mixed deformation, and a logarithmic deformation. Other examples of boundary functions are listed in \cite{Babaei-Aghbolagh:2025hlm,Babaei-Aghbolagh:2025uoz}. Note here that we have adopted some physical conditions for $\ell(\tau)$ implicitly. For example, the signature of the square root is taken so as to reproduce the undeformed limit. It would be possible to formulate the physical conditions like this in terms of $\ell(\tau)$ by following the work \cite{Russo:2024llm} discussed in the context of the CH construction of self-duality invariant theories of non-linear electrodynamics.

\medskip 

Although we have concentrated on integrable deformations of 2D PCM in this paper, the analysis here can easily be generalized to other models by changing the meromorphic function and boundary conditions. For example, even if the seed theory is the same as 2D PCM, homogeneous Yang-Baxter deformations of it \cite{Matsumoto:2015jja} can be accommodated by changing boundary conditions \cite{Delduc:2019whp}. By changing the meromorphic function, one may also consider other seed theories such as 2D PCM with the WZ term and the $\eta$-deformed PCM \cite{Delduc:2013fga}. Therefore, in the framework of the 4D CS theory, it is possible to combine other kinds of integrable deformations with the deformations described by the CH construction. This is an advantage of our result. 

\medskip 

As mentioned above, a lot of generalizations are possible. It would be significant to check whether the corrected term, which is shown to be the trace of the energy-momentum tensor at least in the present 2D PCM case, should be universal or not. To resolve this issue, it is important to study other seed theories and boundary conditions. We will report the result in another place \cite{Fukushima:2026gan,FMY}. It is also interesting to extend our result to the symmetric coset case. The treatment of the grading structure in the 4D CS theory is formulated in \cite{Fukushima:2020dcp}. 

\medskip 

A possible approach to seek for the origin of the corrected term is to uplift the present discussion to the 6D holomorphic CS theory \cite{costello2020topological,Bittleston:2020hfv}. Since the 4D CS theory can be obtained via symmetry reductions, it may be possible to derive the corrected term from the 6D holomorphic CS theory. On the other hand, the 4D Wess-Zumio-Witten (WZW) model can be derived from the 6D holomorphic CS theory. Then, since the 4D WZW model is closely related to the 4D CS theory \cite{Bittleston:2020hfv}, the CH construction may be possible in the 4D WZW model. It is a very interesting future problem. 

\medskip 

There are some potential applications as well. One of the most interesting ones would be to consider the CH construction in the string theory on the AdS$_3\times$S$^3$ background. Both of AdS$_3$ and S$^3$ can be discussed as the target space of 2D PCM. One may consider a couple of the CH constructions for AdS$_3$ and S$^3$\,, but these two should be related through the Virasoro constraints. It would be nice to look for integrable deformations that preserve the consistency of String Theory and study the CFT interpretation of such deformations.

\subsection*{Acknowledgments}

The work of O.\,F.\ was supported by RIKEN Special Postdoctoral Researchers Program and JSPS Grant-in-Aid
for Research Activity Start-up No.\,24K22890.
The work of K.Y. was supported by MEXT KAKENHI Grant-in-Aid for Transformative Research Areas A “Machine Learning Physics” No.\,22H05115, and JSPS Grant-in-Aid for Scientific Research (B) No.\,22H01217 and (C) No.\,25K07313. 

\appendix
\section*{Appendix}

\section{$T\bar{T}$-operator and root $T\bar{T}$-operator}
\label{appB}

Following \cite{Ferko:2022cix,Babaei-Aghbolagh:2025uoz,Babaei-Aghbolagh:2025hlm}, we provide some quantities including the $T\bar{T}$-operator and the root $T\bar{T}$-operator. Now the Lagrangian is given by  $\hat{\mathcal{L}}=\mathcal{F}(x_1,x_2)$, where $x_1$ and $x_2$ are defined in \eqref{x-variable} and $\mathcal{F}$ is going to be determined by solving the PDE \eqref{constraint}.

\medskip 

The energy-momentum tensor defined in \eqref{em-tensor} is written as
\als
{
	\hat T_{\mu\nu}
	&=
    - 2 (\partial_1 \mathcal{F})
    \, \mathrm{tr}\,j_\mu j_\nu
	- 4 (\partial_2 \mathcal{F})
    \, \mathrm{tr}\,j_\mu j_\rho
    \, \mathrm{tr}\,j^\rho j_\nu
	+ \eta_{\mu\nu} \mathcal{F}.
}
From this expression and the the trace identities \eqref{trace-id}, we obtain 
\al
{
    \label{trace-EMtensor}
    \frac{1}{2} \hat T^\mu{}_\mu
	&= - x_1 \partial_1 \mathcal{F} - 2 x_2 \partial_2 \mathcal{F} + \mathcal{F}\,,\\
    \frac{1}{2} \hat T^{\mu\nu} \hat T_{\mu\nu}
	&= 2 x_2 (\partial_1 \mathcal{F})^2
    + (8x_1^2x_2 - 4x_1^4 + 4x_2^2)(\partial_2 \mathcal{F})^2\,,\\
	&\quad
    + (12x_1x_2 - 4x_1^3)(\partial_1 \mathcal{F})(\partial_2 \mathcal{F})
	- 2 \mathcal{F} (x_1\partial_1 \mathcal{F} + 2x_2 \partial_2 \mathcal{F}) + \mathcal{F}^2
	\nonumber\,.
}

\medskip 

By using the above energy-momentum tensor, let us introduce the $T\bar T$ operator $\hat O$ and the root $T\bar T$ operator $\hat R$ defined as
\al
{
    &
    \label{TTbar-op}
	\hat O
	\equiv - \det \hat T_{\mu\nu}
    = \frac{1}{2}\hat T^{\mu\nu}\hat T_{\mu\nu} - \frac{1}{2} (\hat T^\mu{}_\mu)^2\,,\\
    &
    \label{root-TTbar-op}
	\hat R
    \equiv \sqrt{\frac{1}{2}\hat T^{\mu\nu}\hat T_{\mu\nu} - \frac{1}{4}(\hat T^\mu{}_\mu)^2}\,. 
}
In terms of $x_1$ and $x_2$\,, the two operators are rewritten as \cite{Ferko:2022cix}
\al
{
	\hat O
	&=
	- 2(x_1^2 - x_2)[(\partial_1\mathcal{F})^2
	+ 2(x_1^2 - x_2)(\partial_2\mathcal{F})^2
	+ 2x_1(\partial_1\mathcal{F})(\partial_2\mathcal{F})]\\
	& \quad
	+ 2\mathcal{F} (x_1\partial_1 \mathcal{F} + 2x_2 \partial_2 \mathcal{F})
	- \mathcal{F}^2\,, \nonumber\\
	\hat R
	&=
	{\textstyle\sqrt{(2x_2 - x_1^2)(\partial_1\mathcal{F} + 2x_1\partial_2\mathcal{F})^2}}\,.
}
It is also helpful to represent the two operators in terms of $u$ and $v$\,, defined in \eqref{uv-variable}. Then we obtain the following expressions: 
\al
{
	\frac{1}{2} \hat T^\mu{}_\mu
	&= - u\partial_u\mathcal{F} - v\partial_v\mathcal{F} + \mathcal{F}\,,\\
	\frac{1}{2} \hat T^{\mu\nu} \hat T_{\mu\nu}
	&= 2u^2(\partial_u\mathcal{F})^2 + 2v^2(\partial_v\mathcal{F})^2
	- 2\mathcal{F}(u\partial_u\mathcal{F} + v\partial_v\mathcal{F}) + \mathcal{F}^2\,.
}
Then these quantities lead to 
\al
{
	\hat O
	&= 2\mathcal{F} (u\partial_u\mathcal{F} + v\partial_v\mathcal{F})
	- 4uv(\partial_u\mathcal{F})(\partial_v\mathcal{F}) - \mathcal{F}^2\,,\\
	\hat R
	&= \sqrt{(u\partial_u\mathcal{F} - v\partial_v\mathcal{F})^2}\,.
}

\medskip

In particular, after solving the PDE and determining the explicit form of $\mathcal{F}$\,, the above quantities are drastically simplified as \cite{Babaei-Aghbolagh:2025uoz,Babaei-Aghbolagh:2025hlm}
\al
{
    \frac{1}{2} \hat T^\mu{}_\mu
	&= - \tau \ell' + \ell\,,\\
    \frac{1}{2} \hat T^{\mu\nu} \hat T_{\mu\nu}
    &= \ell^2 - 2\tau \ell \ell' + 2\tau^2\ell'^2
}
and the two operators are given by 
\al
{
	\hat O &= -\ell(\ell - 2\tau\ell')\,,\\
	\hat R &= \sqrt{(\tau \ell')^2}\,.
}

\section{A general solution for the root $T\bar T$-flow equation}\label{app:root-TT}

We start with a solution $\mathcal{F}_0$\,, which is represented by a function $\ell_0(\tau)$\,, for the CH problem. Let us consider the root $T\bar T$ flow equation,
\als
{
	\label{root-TTbar-flow}
	\partial_\gamma\mathcal{F}
	= \hat R = \tau\ell'\,,
}
with the flow parameter $\gamma$ and the initial condition $\mathcal{F}(\gamma=0)=\mathcal{F}_0$. Here we assume $\tau\ell'>0$ for simplicity.

\medskip

If $\ell$ depends on $\gamma$, then $\tau$ may also depend on $\gamma$ implicitly through the second equation in \eqref{CH-solution}. As a result, $\ell$ may depend on $\gamma$ both explicitly and implicitly as
\als
{
	\ell=\ell(\tau(u,v,\gamma),\gamma)\,.
}
Then, differentiating the first equation in \eqref{CH-solution} with respect to $\gamma$ gives rise to 
\als
{
	\label{CH-solution-gamma-derivative}
	\partial_\gamma\mathcal{F}
	= \partial_\gamma\ell + \ell'\partial_\gamma\tau
	+ \frac{2u}{\ell'^2}(\partial_\gamma\ell' + \ell''\partial_\gamma\tau)\,,
}
where the second and fourth terms on the right-hand side appear because $\tau$ depends on $\gamma$ like  $\tau=\tau(u,v,\gamma)$. On the other hand, differentiating the second one in 
\eqref{CH-solution}
with respect to $\gamma$ leads to 
\als
{
	\partial_\gamma\tau
	= - \frac{2u}{\ell'^3}(\partial_\gamma\ell' + \ell''\partial_\gamma\tau)\,. \label{B4}
}
Substituting the relation (\ref{B4}) into \eqref{CH-solution-gamma-derivative}, we obtain
\als
{
	\label{CH-solution-gamma-derivative-2}
	\partial_\gamma\mathcal{F}
	= \partial_\gamma\ell\,.
}

\medskip

Combining \eqref{CH-solution-gamma-derivative-2} with the flow equation \eqref{root-TTbar-flow}, we find that 
\als
{
	\partial_\gamma\ell
	= \tau\ell'.
}
Here $\partial_\gamma\ell$ denotes differentiation with respect to the explicit dependence on $\gamma$ only. Solving this equation with the initial condition $\ell(\gamma=0)=\ell_0$, we obtain
\als
{
	\label{conclusion}
	\ell = \ell_0(e^{\gamma}\tau).
}
This is a general solution to the flow equation \eqref{root-TTbar-flow}, which is consistent with the scaling argument of the root $T\bar T$ flow given in \cite{Babaei-Aghbolagh:2025uoz,Babaei-Aghbolagh:2025hlm}. The solution  \eqref{rootTTbar-PCM} is a special case $\ell_0(\tau)=\tau$. The solution (\ref{4.16}) can also be explained by this rescaled form (\ref{conclusion}).

\normalem

\bibliographystyle{utphys}
\bibliography{auxiliary}


\end{document}